\documentclass[a4paper,aip,apl,amsmath,amssymb,reprint,superscriptaddress]{revtex4-1}
\usepackage{graphicx}
\usepackage{amsmath}
\usepackage{dcolumn}
\usepackage{bm}
\usepackage{bbold}
\usepackage{color}
\usepackage{amsfonts}
\usepackage{amssymb}
\usepackage{mathrsfs}
\usepackage{tabularx}
\usepackage{braket}
\usepackage{mathtools}
\usepackage{soul}			

\usepackage{caption}
\usepackage{subcaption}

\begin{document} 
\title{Cavity Magnon Polaritons with Lithium Ferrite and 3D Microwave Resonators at milli-Kelvin Temperatures}
\author{Maxim Goryachev}
\affiliation{ARC Centre of Excellence for Engineered Quantum Systems, School of Physics, University of Western Australia, 35 Stirling Highway, Crawley WA 6009, Australia}

\author{Stuart Watt}
\affiliation{Magnetisation Dynamics and Spintronics Group, School of Physics, University of Western Australia, 35 Stirling Highway, Crawley WA 6009, Australia}

\author{Jeremy Bourhill }
\affiliation{ARC Centre of Excellence for Engineered Quantum Systems, School of Physics, University of Western Australia, 35 Stirling Highway, Crawley WA 6009, Australia}

\author{Mikhail Kostylev}
\affiliation{Magnetisation Dynamics and Spintronics Group, School of Physics, University of Western Australia, 35 Stirling Highway, Crawley WA 6009, Australia}

\author{Michael E. Tobar}
\email{michael.tobar@uwa.edu.au}
\affiliation{ARC Centre of Excellence for Engineered Quantum Systems, School of Physics, University of Western Australia, 35 Stirling Highway, Crawley WA 6009, Australia}

\date{\today}


\begin{abstract}

Single crystal Lithium Ferrite (LiFe) spheres of sub-mm dimension are examined at mK temperatures, microwave frequencies and variable DC magnetic field, for use in hybrid quantum systems and condensed matter and fundamental physics experiments. Strong coupling regimes of the photon-magnon interaction (cavity magnon polariton quasi-particles) were observed with coupling strength of up to 250 MHz at 9.5 GHz (2.6\%) with magnon linewidths of order 4 MHz (with potential improvement to sub-MHz values). We show that the photon-magnon coupling can be significantly improved and exceed that of the widely used Yttrium Iron Garnet crystal, due to the small unit cell of LiFe, allowing twice more spins per unit volume. Magnon mode softening was observed at low DC fields and combined with the normal Zeeman effect creates magnon spin wave modes that are insensitive to first order order magnetic field fluctuations. This effect is observed in the Kittel mode at 5.5 GHz (and another higher order mode at 6.5 GHz) with a DC magnetic field close to 0.19 Tesla.  We show that if the cavity is tuned close to this frequency, the magnon polariton particles exhibit an enhanced range of strong coupling and insensitivity to magnetic field fluctuations with both first order and second order insensitivity to magnetic field as a function of frequency (double magic point clock transition), which could potentially be exploited in cavity QED experiments.

\end{abstract}

\maketitle

\section{Introduction}

Recently, applications of low loss magnonic systems in physics have drawn considerable attention. Modern experimental demonstrations range from Quantum Electrodynamics (QED)\cite{Goryachev:2014ab,Tabuchi:2014aa} to fundamental physics\cite{sikivie,ferri-axions,Barbieri:2017aa}. The main focus of all these experiments is on the rare-earth iron garnets, particularly, Yttrium Iron Garnet (YIG), a material that combines good photon and magnon properties over the entire microwave spectrum as well as the possibility to enhance its performance for particular applications with chemical substitution. For room temperature applications, it has been widely used as a tunable element in many common microwave devices such as filters and oscillators\cite{6243735}, as a material for nonlinear studies\cite{McKinstry:1985aa,Wang:2016aa}, and as a platform for prospective magnon based information processing devices\cite{Chumak:2014aa,Chumak:2015aa}. Additionally, YIG has been considered as a material of choice for many implementations of hybrid quantum systems\cite{Soykal:2010ly,Tabuchi:2015aa,Zhang:2014aa,Bai:2015aa,Klingler:2016aa,Morris:2017aa,Andrich:2017aa,Lachance-Quirion:2017aa}. On the other hand, it suffers from high losses in the optical frequency band as well as very high refractive index, which makes optical coupling inefficient and very hard to implement\cite{Zhang:2016aa}. Also, the performance of YIG for many applications may suffer from coupling to higher order magnon modes due to possible crystal imperfections, especially at very low temperatures. Thus, it is interesting to compare this material to similar ferrimagnetic single crystals that have very high spin density and narrow magnon linewidths. 

One particular candidate is the single-phase crystal Lithium Ferrite (LiFe) of empirical formula Li$_{0.5}$Fe$_{2.5}$O$_4$\cite{Mazen:2015aa,Soreto:2017aa,Teixeira:2012aa,1128308,1067269}, which belongs to the cubic spinel ferrites family with properties competitive with YIG\cite{1128308,1067269}.  LiFe belongs to the class of soft magnets with square magnetic hysteresis loops, large magnetization and strong anisotropy. Some additional properties such as strong Dzyaloshinskii-Moriya interaction lead to proposals for novel magnonic devices based on relativistic band engineering of LiFe crystals\cite{PhysRevB.92.184419}. Besides microwave applications, the material is used for coatings of anodes for lithium ion batteries\cite{Zeng:2014aa}, Terrahertz devices\cite{Kinoshita:2016aa} and gas sensing\cite{Patil:2016aa}. 

Experiments that couple magnon spin wave modes to photonic cavities to create cavity magnon polariton (CMP) two-level systems have recently been a fertile area of investigation. For example CMP bistability \cite{PhysRevLett.120.057202}, CMP exceptional points \cite{Zhang:2017qf} and CMP manipulation of distant spin currents \cite{PhysRevLett.118.217201} have recently been observed. In this work, we couple a 3D lumped photonic resonant cavity to a LiFe sphere to create CMPs and illustrate the potential for QED at microwave frequencies and mK temperatures. Furthermore, we observe in the dispersive regime of the cavity, magnon spin-wave modes, which are first order insensitive to magnetic field close to 5.5 GHz in frequency. We show that if the cavity mode can be tuned to this magnetic field insensitive point of the magnon, where $df/dB=0$, we create a two-mode CMP with enhanced coupling range and reduced curvature (point of inflection, with both $df_\text{CMP}/dB=0$ and $d^2f_\text{CMP}/dB^2=0$), similar to a double magic point atomic clock transition\cite{PhysRevA.90.053416}. Previously, solid state clock transitions have been observed in nuclear, electron and hyperfine solid state spin systems, such as NV in diamond\cite{PhysRevA.87.032118,Dolde:2011ul}, Eu$^{3+}$ in YSO\cite{Zhong:2015gf}, Bismuth in Si\cite{Wolfowicz:2013pd,PhysRevB.89.155202,PhysRevB.92.161403,PhysRevB.86.245301,PhysRevB.85.094404} as well as molecular spin systems\cite{Shiddiq:2016mz,PhysRevLett.119.140801}.

\section{Cavity Magnon Polaritons with a Lithium Ferrite Sphere}

The LiFe specimens were highly polished $0.46$~mm diameter spheres glued onto a microwave ceramic post with specific orientation. The post as well with the external DC field are oriented along the (110) crystallographic direction of the sphere.The spheres exhibit ferrimagnetic resonance linewidth $\Delta H$ of 1~Oe and the saturation magnetization $\mu_0 M_s$ of $0.37$~T at room temperature. The refractive index of LiFe thin crystal is measured to be $\sim 2.6$ at 800 nm and $\sim 2.3$ at 1550 nm\cite{Kinoshita:2016aa}, and permittivity at microwave frequencies is approximately 15, close to that of YIG\cite{Kishan:1994aa,PhysRevB.93.144420}. In this work, we investigate LiFe properties using cavity methods at 20 mK and very low photon excitation number. For this purpose, a cavity containing a LiFe crystal is attached to the lowest temperature stage of a Dilution Refrigerator inside a superconducting magnet and characterised as a function of the DC magnetic field using continuous wave excitation. The experimental setup is described in detail in previous works\cite{PhysRevB.88.224426,Goryachev:2014aa}.

\begin{figure}[h!]
     \begin{center}
            \includegraphics[width=0.45\textwidth]{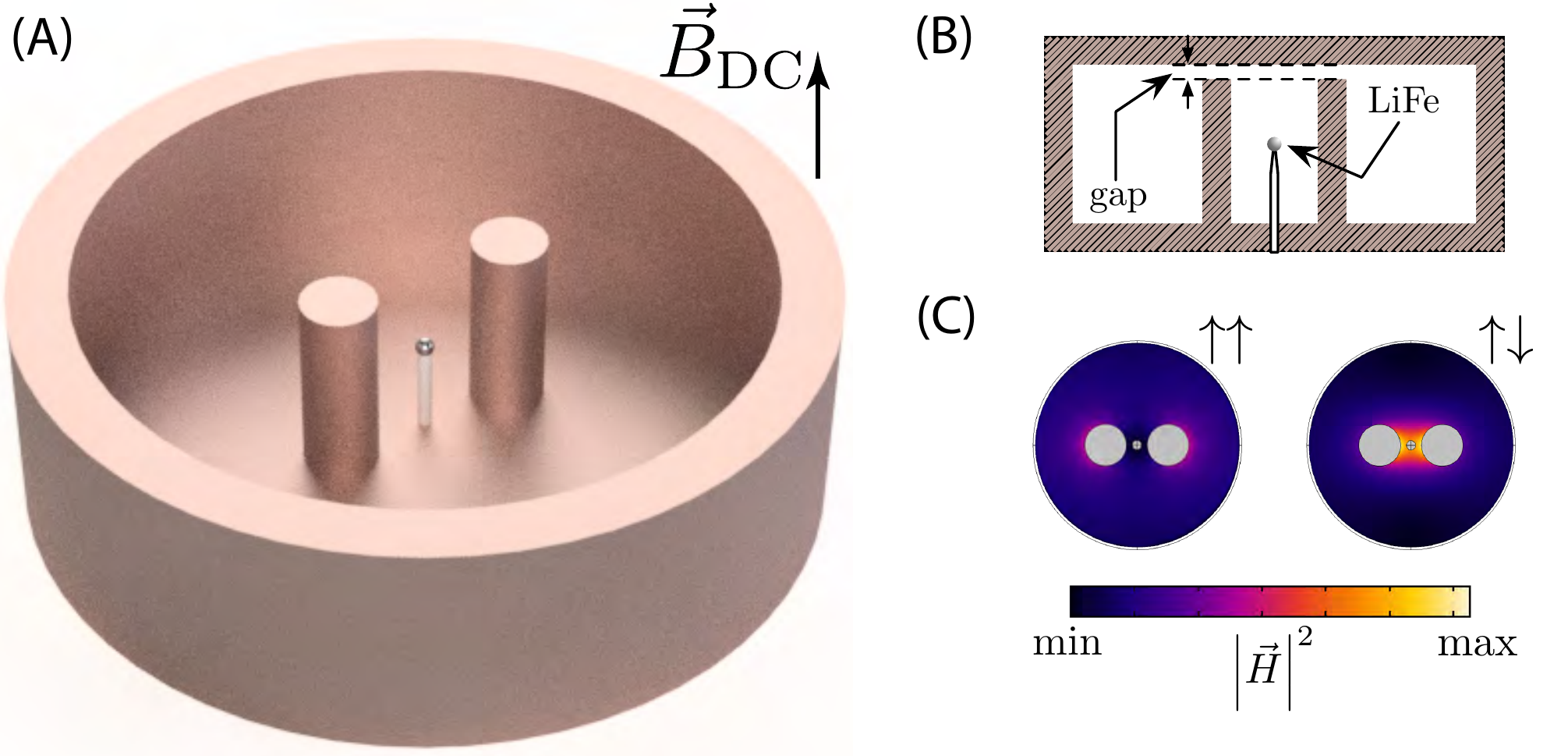}
            \end{center}
    \caption{(A) Double post reentrant cavity (without top lid) with a LiFe sphere in external magnetic field, with cross-section in (B). This cavity supports a dark ${\uparrow\uparrow}$ and a bright mode ${\downarrow\uparrow}$ as shown in (C). }%
   \label{cavity}
\end{figure}

The material is characterised using a re-entrant microwave cavity\cite{reen1,reen0}. Such resonators are typically cylindric, with a metallic post attached to one of the conducting faces which stops just short of the opposite face, forming a small gap. The re-entrant mode has the electrical field confined in this gap and the magnetic field around the post, and thus the metallic rod forms a 3D lumped element LC resonator. The re-entrant mode resonance frequency can be calculated as a resonance of a corresponding LC-circuit, thus the resonance frequency is proportional to a square root of the gap distance\cite{reen1,reen0}. In the double-post structure as shown in Fig.~\ref{cavity}, low and high frequency re-entrant resonances have symmetric (co-directional currents and electric fields for both posts) and antisymmetric (contra-directional currents and electric fields for both posts) structure respectively. As a result, the symmetric resonance expels magnetic field from the space between the posts and is henceforth referred to as the dark mode, whilst the anti-symmetric mode focuses the field and will be referred to as the bright mode. This property has been used to enhance the spin-photon coupling by placing crystal samples of sub-wavelength size between the posts\cite{Goryachev:2014aa,Creedon:2015aa}. The position of the crystal inside the cavity and orientation of the external field is shown in Fig.~\ref{cavity}. The other advantage of this method is the ability to spatially separate the magnetic properties of test crystals from electric ones. This property is achieved due to the fact that for the re-entrant type structure, most of the electric field is concentrated in the gap, while the magnetic field spreads around the posts leading to their strong spatial separation. For this reason, dielectric properties of LiFe crystals need not be considered. 

\begin{figure}[h!]
     \begin{center}
            \includegraphics[width=0.5\textwidth]{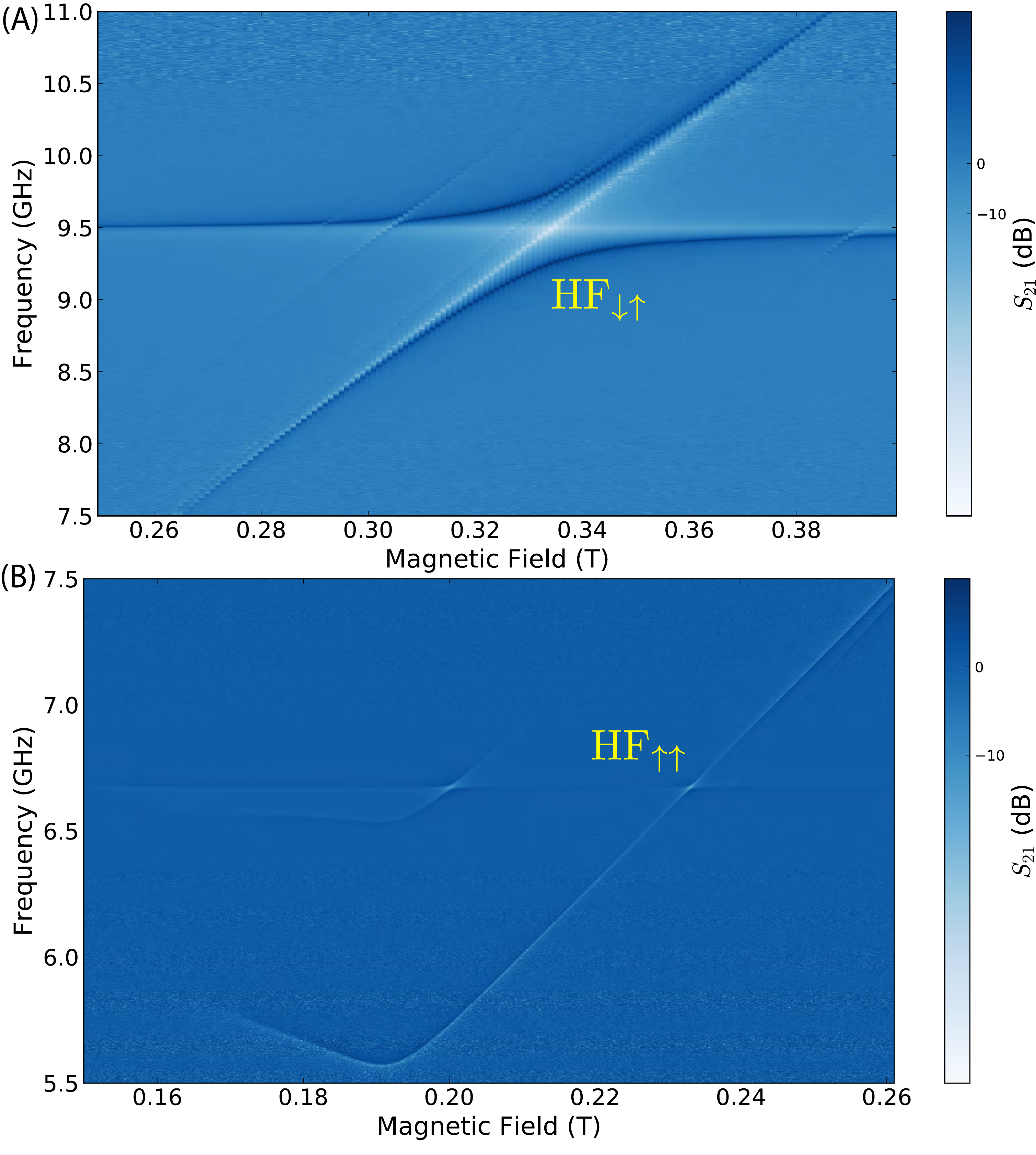}
            \end{center}
    \caption{Frequency response of the LiFe sphere-cavity system as a function of external DC field for the HF case: (A) close to the bright mode ${\downarrow\uparrow}$, (B) near dark mode ${\uparrow\uparrow}$.}%
   \label{high}
\end{figure}

\begin{figure}[h!]
     \begin{center}
            \includegraphics[width=0.5\textwidth]{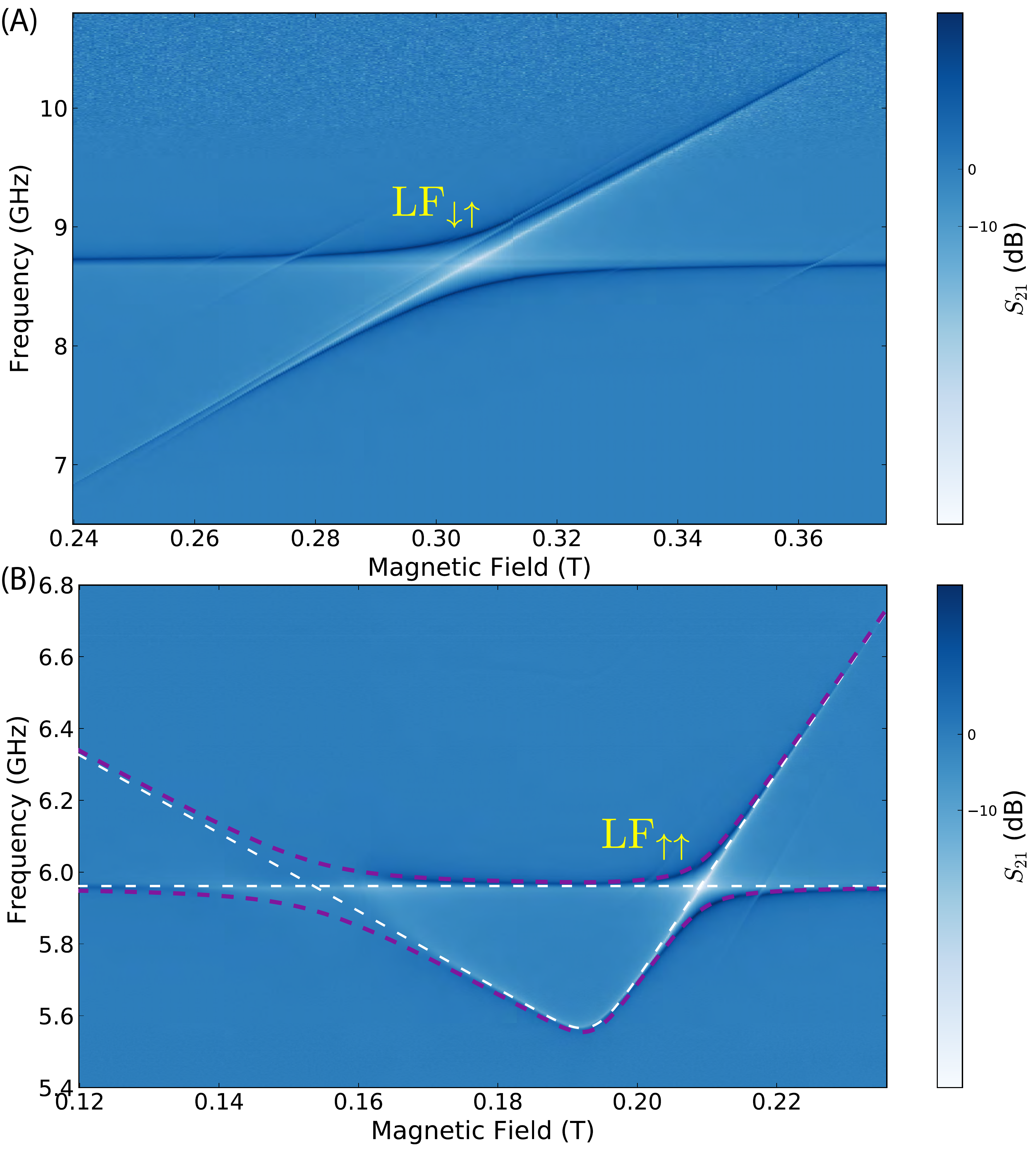}
            \end{center}
    \caption{Frequency response of the LiFe sphere-cavity system as a function of external DC field for the LF case: (A) close to the bright mode, (B) close to the dark mode ${\uparrow\uparrow}$, which has been fitted with a two-mode model, with the coupled (uncoupled) solutions in dashed purple (white) and shows two mode crossings. The lower crossing in (B) is not in the strongly coupled regime due to excess magnon losses at low magnetic field, while the upper crossing is strongly coupled.}%
   \label{low}
\end{figure}

The cavity was measured at temperatures near 20~mK in a dilution refrigerator as a function of DC magnetic using a superconducting magnet. The excitation signals were attenuated at 4~K and 20~mK stages down to the level of a few photons. The transmitted photons were amplified by a low noise 4~K amplifier separated from the cavity by a circulator. The cavity of 10~mm diameter and 3.8~mm height was made of oxygen free Copper with 2~mm diameter posts separated by 1.5mm from the cavity centre. With these dimensions, the filling factors (ratio of the magnetic energy stored in the LiFe sphere and the total magnetic energy in the cavity) for the bright and dark mode were estimated to be 0.1\% and $10^{-4}$\% respectively. The former may be greatly increased by minimising the cavity height and separation between the posts\cite{Goryachev:2014aa} without change of resonant frequency. 

The photon-magnon coupling experiment with the $0.46$~mm diameter sphere was repeated twice using different values of the gap between the top of the posts and the cavity lid. The system responses for the cases of larger gaps and smaller gap as a function of external magnetic field and frequency are shown in Fig.~\ref{high} and \ref{low} respectively. A run with larger gaps gives higher resonance frequencies and is referred to as Higher Frequency (HF) experiment, accordingly, the smaller gap case results in Lower Frequencies (LF). In each case, the system is characterised in the vicinity of the dark and bright modes with corresponding frequencies $f_{\uparrow\uparrow}$ and $f_{\downarrow\uparrow}$. 

Near each avoided level crossing the system may be approximated by the simplest linearly coupled two harmonic oscillator model: 
\begin{multline}
\begin{aligned}
	\label{hamiltonian1}
	\displaystyle  H/h = \nu_\text{ph}a^\dagger a + \nu_m(B)b^\dagger b + g (a^\dagger b + b^\dagger a),
\end{aligned}
\end{multline}
where $a^\dagger$ ($a$) and $b^\dagger$ ($b$) are creation (annihilation) operators for photonic and magnon modes with corresponding  resonance frequencies of $\nu_\text{ph}$ and $\nu_m(B)$, and the magnon resonance frequency is controlled via the external magnetic field ($B$) due to the Zeeman effect $\nu_m = \text{g} \mu_B (B +B_0)/\hbar$.
Here $\mu_B$ is the Bohr magneton, g is the effective Land{\'e} g-factor and $B_0$ is an effective magnetic field bias. The summary of the model parameters for two modes of two experiments are given in Table~\ref{table1}. For all experiments with the 0.46~mm (110) sample, the magnon linewidth is estimated to be on the order of 9~MHz. The maximum observed coupling is $250$~MHz at 9.5~GHz (i.e. $g/\omega = 2.6$\%), which would be improved further by increasing the filling factor. If the filling factor could be increased to even 3\%, a value which was previously demonstrated with a YIG sphere inside the same type of cavity\cite{Goryachev:2014aa}, one could increase the coupling to 1.3 GHz or 13.7\% of the resonant frequency. In the case of the YIG sphere experiment, magnon-photon coupling of 2~GHz at 20.6~GHz or 9.7\%\cite{Goryachev:2014aa} was measured, and was achieved using a sphere of 0.8 mm diameter hence a 5.3 times larger volume than the LiFe described here.


\section{Magnetisation Properties of Lithium Ferrite and Comparison to YIG}

For several particular applications, the spin density or the total number of spins is the important parameter for comparison. Thus, it is interesting to compare LiFe to the most popular magnonic material YIG. Unlike YIG, LiFe exhibits spinel structure with one Fe$^{3+}$ ion occupying the tetrahedral sites compared to 1.5 ions occupying the octahedral sites and oriented antiparallel, which means there are, $\Delta n=0.5$, dominant octahedral ions in each unit cell. Both types are Fe$^{3+}$, which carry 5 Bohr magnetons and thus determines the effective spin density to be $N_\textrm{eff} = {M_s}/{\mu_B} \approx 3.2 \times 10^{22} \textrm{~cm}^{-3}$.
With a magnetic moment of a unit cell equal to 10$\mu_B$ and a typical $M_s$ value of $0.175$~T, the effective spin density of YIG may be estimated to be $1.5\times 10^{22}$~cm$^{-3}$, two times less. Compared to YIG, the larger spin density of LiFe is due its unit cell occupying a smaller volume, which is verified via the considerably stronger normalised photon-magnon coupling as discussed above.

Besides the main uniform magnetisation precession mode, the sphere under cryogenic conditions exhibit a number of higher order magnon modes as seen for both bright and dark modes in both Fig.~\ref{high} and \ref{low}. These effects are also seen for all tests of YIG at low temperatures and typically attributed to arising anisotropies. It is found that the frequencies of these modes cannot be predicted by the same theory\cite{fletcher} as successfully applied to YIG spheres\cite{Goryachev:2014aa} due to the strong anisotropy.




\begin{table}[]
\centering
\caption{Parameters of avoided level crossing between the uniform magnon made to cavity resonances.}
\label{table1}
\begin{tabular}{c|c|c|c|c|c}
\hline
Mode & $\nu_\text{ph}$ (GHz)  & g & $B_0$ (mT) & $g$ (MHz) & \\ \hline
HF$_{\uparrow\uparrow}$     & 6.67 & $\sim2$ & 6.4 & 9 \\
HF$_{\downarrow\uparrow}$     & 9.5 & $\sim2$ & 10 & 250 \\
LF$_{\uparrow\uparrow}$     &  5.96 & $\sim2$ & 8.2 & 68\\
LF$_{\downarrow\uparrow}$     &  8.7 & $\sim2$ & 9 & 240 \\ \hline
\end{tabular}
\end{table}

\begin{figure}[h!]
     \begin{center}
            \includegraphics[width=0.5\textwidth]{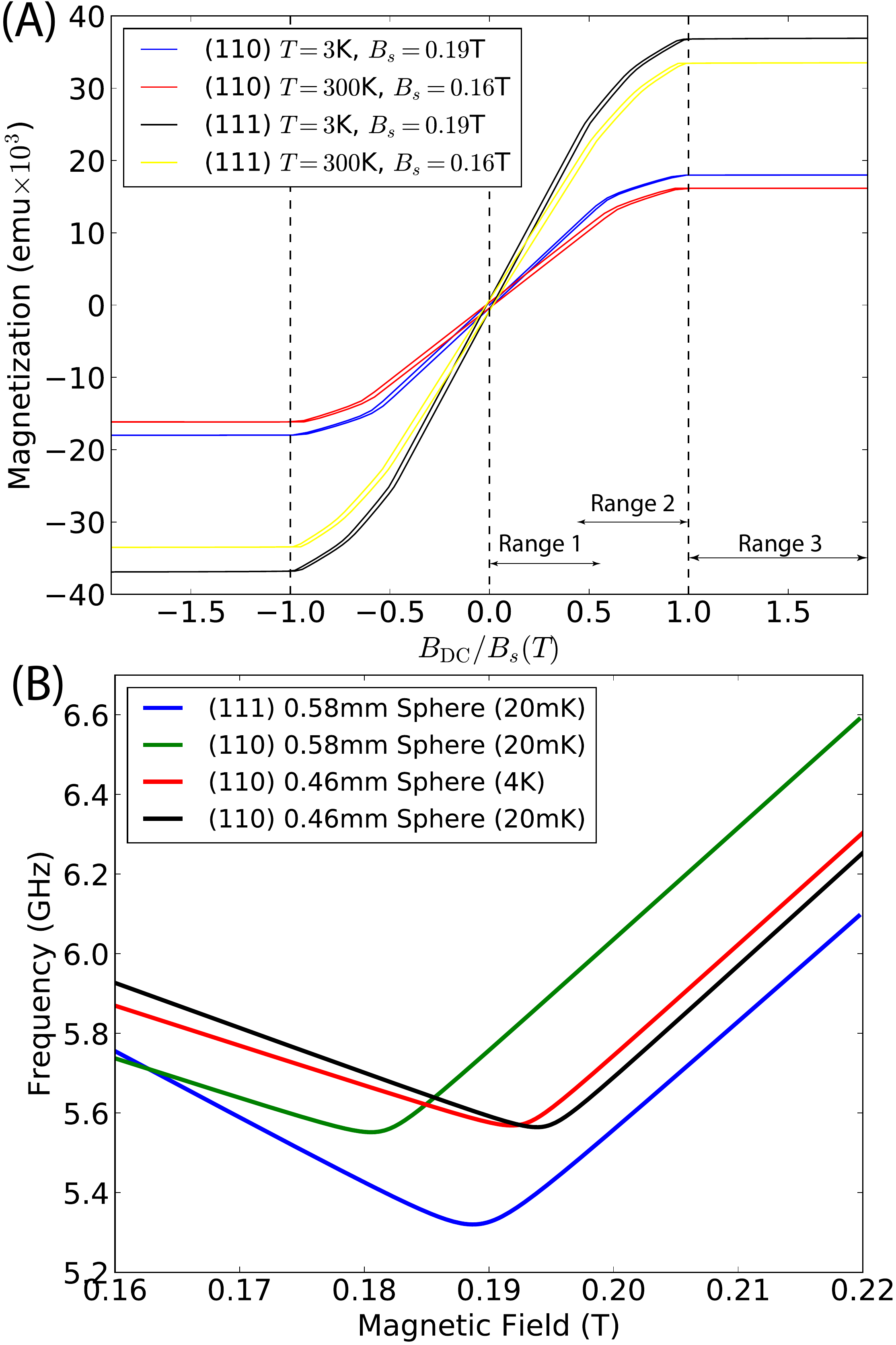}
            \end{center}
    \caption{(A) Magnetisation loops for LeFe spheres at 300 K and 3 K with the external field scaled to the saturation field. (B) Curve fits to resonance peaks at low frequencies.}%
   \label{loops}
\end{figure}

Here one has to mention that there has been only one recent experiment where FMR measurements for LiFe were taken over a broad range of frequencies\cite{Pachauri:2015aa}. Therefore, not much detail can be found in the literature on the field dependence of the FMR frequency for this type of ferromagnetic spinel. What is known, is that our LiFe material should possess a cubic magneto-crystalline anisotropy\cite{Greifer:1969aa,Krupicka} with (111) being an easy magnetisation axis (and (100) being the axis along which the resonance linewidth is minimal\cite{Denton:1962aa}). However, the shapes of the hysteresis loops for the spheres (Fig.~\ref{loops} (A)) taken for (110) and (111) directions are qualitatively the same, and both are characteristic for hysteresis measurements taken at an angle with an easy axis. This may suggest that in addition to the magnetocrystalline cubic anisotropy, our samples may also possess a uni-axial anisotropy potentially induced during sphere grinding, as the axis of this anisotropy is not along (110) or (111). 
A more detailed experimental analysis of the nature of the spheres magnetic anisotropy is beyond the scope of this paper. Importantly, we show there is enough evidence that the spheres possess a significant anisotropy field. This evidence is not only the shape of the hysteresis loops, but also a fact that linear fits to sections of the lines in Fig.~\ref{loops} (B) which correspond to the magnetically saturated state of the spheres deliver an vertical offset of $f_b = 440$~MHz from zero field for the (110) 0.58mm sphere at 20 mK (see Table~\ref{table2} for other cases). 

Given this evidence, the non-monotonicity of the curves in Fig.~\ref{high} (B) and Fig.~\ref{low} (B) can be explained as mode softening\cite{Silber:1967aa}. This is seen in their magnetic field dependence resulting in a magnon magnetic field insensitivity ($dB/df = 0$) at $B_s = 0.191$~T as seen in as turn-over points in the magnon resonance dependence on the external field. This value of the turn-over point is associated with the properties of the hysteresis loop of LiFe. To identify the softening field, the sample hysteresis loop was measured at 3 K as well as at room temperature and plotted in Fig.~\ref{loops} (A). As it is seen from this figure, the magnetic hysteresis loop measurement demonstrates that $B_s$ corresponds to the saturating field of the material. 

In Silber et al\cite{Silber:1967aa}, it is shown that if a sphere is magnetized at an angle to the easy axis, the frequency vs field dependence can be separated into three sections. Above the saturating  field, all spins in the material are perfectly co-aligned and also co-aligned with the external  field. This results in dynamics obeying Kittel equation for the ferromagnetic resonance frequency (Field Range 3). Below this field, two regimes can be identified. The first regime is in the closest vicinity of the saturating field (Field Range 2). Here the material is in a single-domain state, as above the saturating field, but the magnetisation vector is not collinear to the applied field - the vector is closer to the easy axis than the field. With an increase in the field, the vector rotates closer to the field. This leads to an increase in the FMR frequency. Importantly, in contrast to the fully saturated state, the dependence is not a straight line, with the curvature diminishing with an approach to the magnetic saturation (at which the static magnetization vector becomes aligned to the field). At smaller fields (Field Range 1), the sphere is broken into magnetic domains with magnetization vectors in them aligned along the easy axis (or multiple easy axes for cubic anisotropy). As shown theoretically by Smith and Beljers\cite{smit1955}, in this regime, the resonance decreases to a finite value at the critical field, and then increases following the single-domain theory from Silber et al.\cite{Silber:1967aa}. 

This explanation is consistent with the variation in the resonance linewidth with the applied field seen in Figs.~\ref{high} (B) and \ref{low} (B). The FMR responses for Ranges 2 and 3 are characterised by a relatively narrow resonance line. This is coherent with the single-domain state of static magnetization for these ranges. For Range 1 the resonance linewidth monotonically broadens as the magnetic field decreases. This is consistent with a more developed domain structure for lower magnetic field values. At room temperatures, parameter $B_s$ was measured to be $0.154$~T giving the expected temperature dependence for this value. Finally, it is also seen from Fig.~\ref{high} (B) that the higher order magnon mode exhibits the same softening phenomenon at the same value of external magnetic field.

To characterise the softening phenomenon, LiFe samples of different size and orientation with respect to the applied field were measured. The results are summarised in table~\ref{table2} and shown in Fig.~\ref{loops} (B), which compares parameters of the magnon resonance frequency dependencies on the external magnetic field. Here g$_-$, g$_+$, $B_s$, $f_s$, $f_b$, $\delta$ are the effective g-factor below the saturation magnetisation, g-factor above the saturation magnetisation, the softening field (identified as  saturation magnetisation), frequency of the magnetic field insensitive point where $df/dB=0$, zero field frequency bias for a linear fit to the magnetically saturated state, and an estimation for the magnon linewidth at $B>B_s$ respectively. Firstly, for the (110) 0.46~mm diameter sphere, the saturation magnetisation field decreases by going from $20$~mK to $4$~K, which is consistent with the room temperature to 20 mK comparison. Secondly, the test reveals a dependence on sphere size, with the turnover point{'}s field value decreasing for larger spheres. Thirdly, the sphere with (111) orientation demonstrates significantly different minimum frequency and effective g-factors. 
From the table it follows that the gyromagnetic ratio (or effective g-factor) depends on orientation, with the value for (110) samples exceeding previously measured values at room temperature\cite{Pachauri:2015aa}.

\begin{table}[]
\centering
\caption{Parameters of the gyromagnetic curve for the uniform precession magnon mode.}
\label{table2}
\begin{tabular}{c|c|c|c|c|c|c}
\hline
Sample & $g_-$ & $g_+$ & $B_s$ & $f_s$ & $f_b$ & $\delta$ \\
&&&(mT)&(GHz)&(MHz)&(MHz)\\ \hline
(110) 0.46mm, 20 mK    & -0.81 & 2.02 & 194 & 5.56 & 90 & 9\\
(110) 0.46mm, 4 K    & -0.72 & 2.01 & 192  & 5.57 & 142 & 9\\
(110) 0.58mm, 20 mK    & -0.71 & 2.01 & 180 & 5.55 & 440 & 4\\
(111) 0.58mm, 20 mK    & -1.19 & 1.96 & 189 & 5.32 & 10 & 5\\
\hline
\end{tabular}
\end{table}

Table~\ref{table2} also compares estimations of magnon linewidths for different samples. It suggests that the bigger samples exhibit some reduction in linewidths with the lowest value of $4$~MHz at $9$~GHz, possibly due to reduction of the contribution of surface magnon scattering to the total linewidth. Nevertheless, this value is larger than the lowest demonstrated linewidths for state-of-the-art YIG spheres at mK temperatures, but within an order of magnitude\cite{Goryachev:2014ab,Tabuchi:2014aa}. 
The observed minimum value of linewidth is in accordance with the mean value of 1.35 Oe at 5.1 GHz (or equivalently 3.8 MHz) for a batch of selected low defect concentration samples measured at 4.2 K\cite{Spencer:1968aa}. To the best of our knowledge, the lowest reported values is 0.274 Oe (equivalently 770 kHz) for a (100) sphere at 4 K around 10.8 GHz\cite{Denton:1962aa}. Such low values of magnon linewidths make this ferrite material a strong alternative to YIG spheres taking into account the fact that LiFe material research has not been developing with the same pace as the YIG industry in the last 50 years.
Also, uniform precession modes for both samples exhibit strong coupling of 50 MHz to the dark cavity mode at around 8 GHz, and the second strongest higher order magnon mode is coupled at the rate of 11 MHz. 

\section{Cavity Magnon Polariton with Reduced Magnetic Field Sensitivity}

\begin{figure}
\centering
\includegraphics[width=\columnwidth]{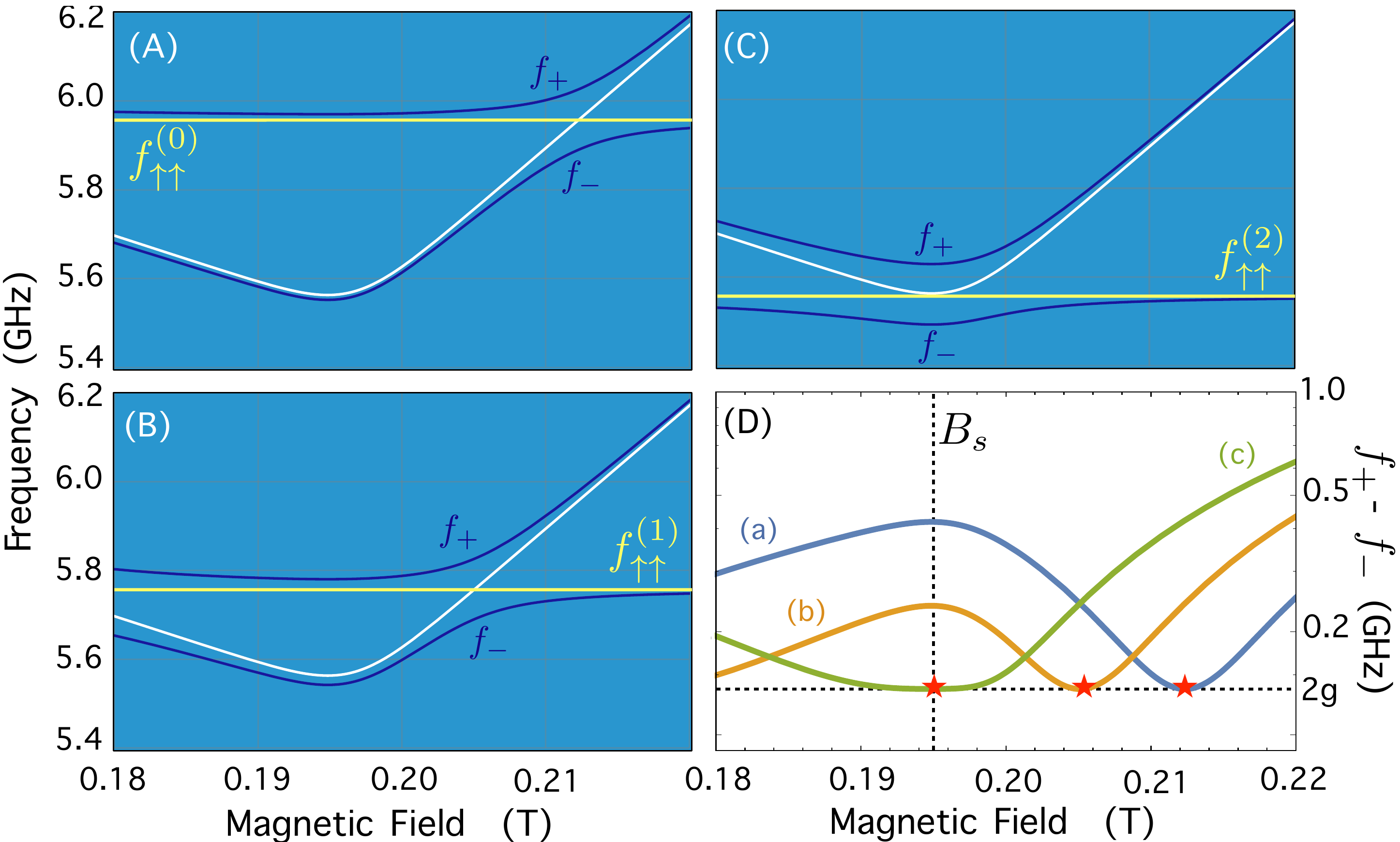}
\caption{Modelled interaction of mode crossing shown in Fig.\ref{low}(B), lowered from its LF$_{\uparrow\uparrow}$ frequency of $f_{\uparrow\uparrow}^{(0)}=5.96$ GHz (A), to $f_{\uparrow\uparrow}^{(1)}=5.76$ GHz (B), and finally $f_{\uparrow\uparrow}^{(2)}=f_s=5.56$ GHz (C). The white and yellow curves represent the uncoupled magnon and cavity modes{'} magnetic field dependence, respectively, whilst the dark lines represent the behaviour when the two are coupled at a rate of $g=68$ MHz. The frequency difference between the higher frequency branch and lower frequency branch (the transition frequency, $f_\text{CMP}=f_+-f_-$) is then plotted for each of these three cases in (D). The red stars represent the location where the photon and magnon maximally hybridise.}
\label{clocktransition}
\end{figure}

Fig. \ref{clocktransition} shows the frequency response between the dark mode and magnons in the (110) 0.46 mm, 20 mK LiFe sample as the cavity photon frequency is tuned to the turning point of the magnon mode at 5.56 GHz. In our experiment the cavity dark mode is only tuned 400 MHz away (Fig.~\ref{clocktransition} (A)) and we show how the characteristics change using a two-mode coupled model, when the cavity frequency is also tuned 200 MHz away (Fig.~\ref{clocktransition} (B)) and then to the point of exact tuning (Fig.~\ref{clocktransition} (C)). When the cavity is tuned to the point of exact tuning, the CMP transition frequency, $f_\text{CMP}=f_{+}-f_{-}$, range is enhanced, resulting in a lower curvature CMP transition frequency versus B field characteristic and a broader range of magnetic field where strong coupling is attained (Fig.~\ref{clocktransition} (D)).

\begin{figure}
\centering
\includegraphics[width=\columnwidth]{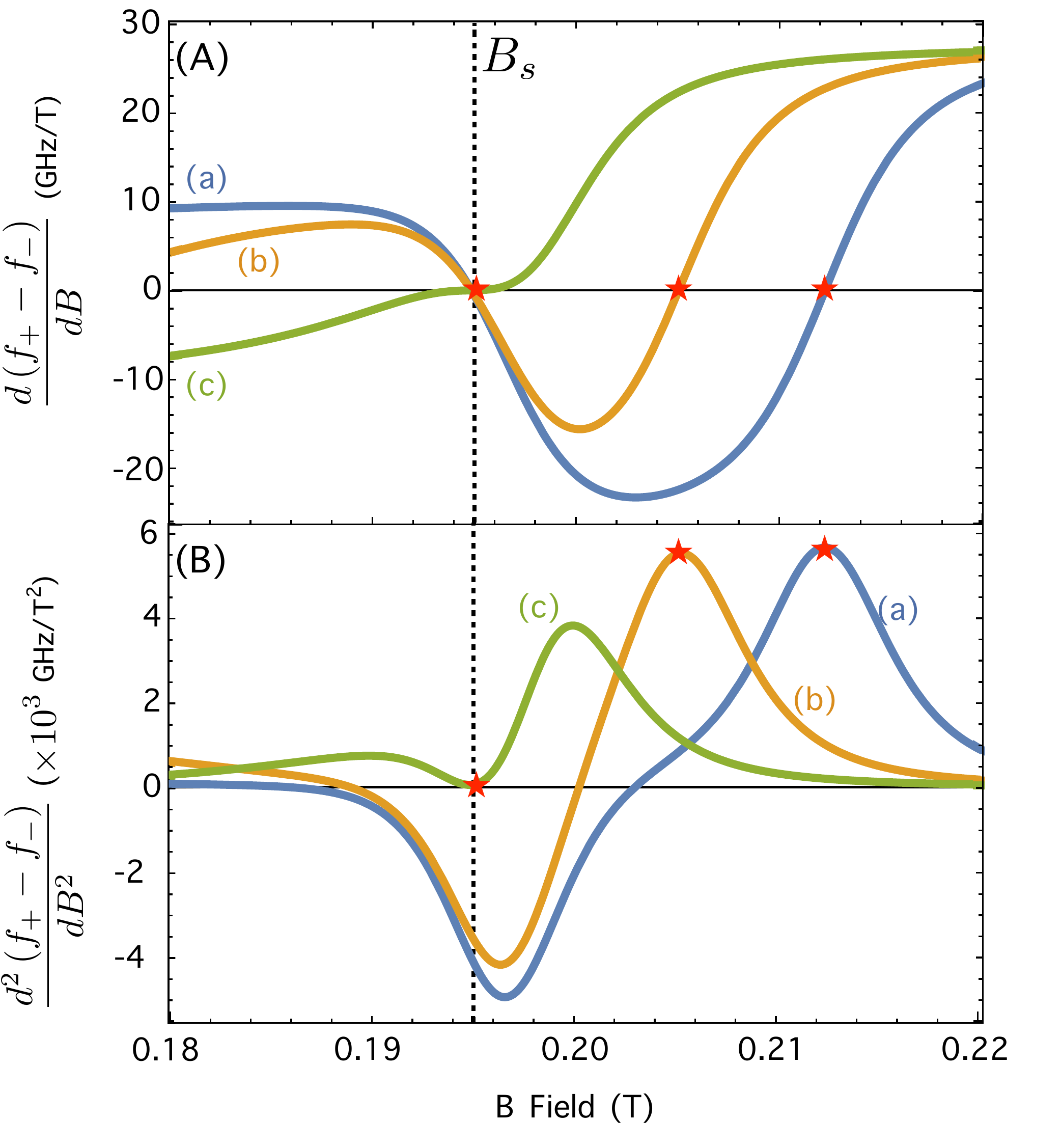}
\caption{First (A) and second (B) derivatives of the frequency of the CMP quasi particle transition with respect to magnetic field for the same values of detuning as shown in Fig.~\ref{clocktransition}.}
\label{inflection}
\end{figure}

To study the properties of the CMP transition as we tune to the magnon mode turn over point, we plot the first and second derivative of $f_\text{CMP}$ with respect to magnetic field as a function of photonic mode cavity tuning, which is illustrated in Fig.\ref{inflection}. The figure clearly shows the point of maximal hybridisation (centre of an avoided level crossing) occurs when $df_\text{CMP}/dB=0$ for all cavity mode de-tunings. However, as the photonic mode cavity frequency approaches the magnon mode turning point the second order is reduced and finally reaches zero at the maximal hybridisation (curve (c)). This is similar to a double magic point atomic clock transition\cite{PhysRevA.90.053416} which significantly reduces the effects of magnetic field biasing fluctuations and could potentially be exploited in cavity QED experiments.

\section{Conclusion}

In conclusion, we investigated the properties of LiFe as a magnonic material coupled to a photonic cavity at ultra-low temperatures. It is demonstrated that due to very high spin density and relatively low magnon losses, LiFe spheres allow strong coupling regimes of magnon-photon interactions similar to YIG (coupling strengths of hundreds of MHz), and to thus create cavity magnon polariton two level systems. With further geometry optimisations, ultra-strong and super-strong coupling regimes are also possible. Although, the LiFe samples in this work are modestly inferior to the best YIG spheres in terms of magnon losses, in principle, this ferrite crystal may demonstrate comparable linewidths. Additionally, LiFe has additional advantages related to enhancing the range of strong coupling and insensitivity to magnetic field fluctuations due to the softening behaviour combining with the normal Zeeman effect. 

This work was supported by the Australian Research Council Grant No. CE170100009.

\end{document}